# Ferroelectric and dielectric properties of $Hf_{0.5}Zr_{0.5}O_2$ thin film near morphotropic phase boundary

Alireza Kashir*, Hyunsang Hwang**

Center for Single Atom–based Semiconductor Device and Department of Materials Science and Engineering, Pohang University of Science and Technology (POSTECH), Pohang, Republic of Korea

*kashir@postech.ac.kr, **hwanghs@postech.ac.kr

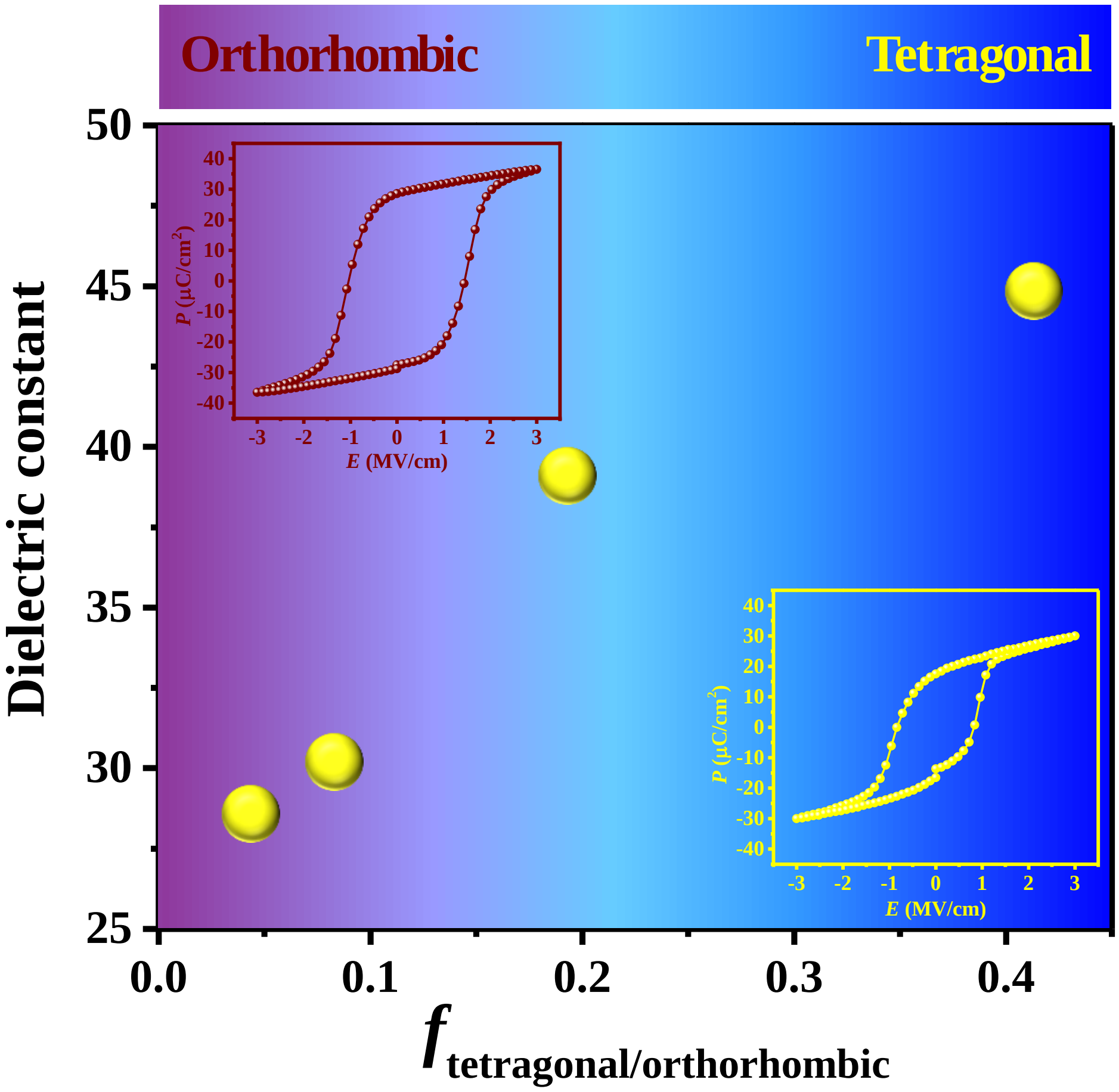

# Abstract

Recently, based on the phase-field modeling, it was predicted that $Hf_{1-x}Zr_xO_2$ (HZO) exhibits the morphotropic phase boundary (MPB) in its compositional phase diagram. Here, we investigate the effect of structural changes between tetragonal (t) and orthorhombic (o) phases on the ferroelectric and dielectric properties of HZO films to probe the existence of MPB region. The structural analysis show that by adjusting the ozone dosage during the atomic layer deposition process and annealing conditions, different ratios of t- to o-phases $\left(f_{\frac{t}{o}}\right)$ were achieved which consequently affect the ferroelectric and dielectric properties of the samples. Polarization versus electric field measurements show a remarkable increase in ferroelectric characteristics ($P_r$ and $E_c$) of the sample that contains the minimum t-phase fraction $\left(f_{\frac{t}{o}} \sim 0.04\right)$. This sample shows the lowest $\epsilon_r$ compared to the other samples which is due to the formation of ferroelectric o-phase. The sample that contains the maximum $f_{\frac{t}{o}} \sim 0.41$ demonstrates the highest dielectric response. By adjusting the $f_{\frac{t}{o}}$, a large $\epsilon_r$ of ~ 55 is achieved. Our study reveals a direct relation between $f_{\frac{t}{o}}$ and $\epsilon_r$ of HZO thin films which can be understood by considering the density of MPB region.

## Introduction

Since the discovery of ferroelectric properties in $HfO_2$-based materials [1], numerous advantages such as simple structures, strong binding energy between oxygen and transition metal ions, large bandgap (~5.3–5.7 eV), and compatibility with current complementary metal–oxide semiconductor technologies led to extensive research that focused on potential diverse applications such as ferroelectric memory, ferroelectric field effect transistors (Fe-FETs), pyroelectric sensors, and energy harvesters. Meanwhile, from the material survey of ferroelectrics, it is conventional that the structural phase transition induced by a compositional change plays a special role in obtaining excellent piezoelectric and dielectric properties [2-9]. The phase boundary known as morphotropic phase boundary (MPB) separates regions of different symmetry by varying their composition in ferroelectrics [2-9]. In fact, when different structural phases are almost degenerate near the MPB, rotations of polarization occur in response to external electric fields instead of a change in the magnitude of polarization [5]. Consequently, materials exhibit maximum piezoelectric and dielectric responses near the MPB as several phases co-exist and there is nearly no energy barrier separating them. The MPB has garnered significant practical interest because the variable that drives the transition (i.e., compositions) is inherent. Therefore, it provides an alternative approach to boost the dielectric constant $\epsilon_r$ without degrading the bandgap. MPB ceramics are the basis of a wide range of piezoelectric and dielectric technologies, such as high-*k* materials, sensors, actuators, smart systems, ultrasound generation and sensing and underwater acoustics. These applications motivated researchers to investigate the existence of MPB in the novel promising ceramic compounds e.g., $PbHfO_3$–$PbTiO_3$–$Pb(Mg_{1/3}Nb_{2/3})O_3$ [10, 11], lead free $BiFeO_3$-$BaTiO_3$ [12, 13], $Pb(In_{1/2}Nb_{1/2})O_3$–$Pb(Mg_{1/3}Nb_{2/3})O_3$–$PbTiO_3$ [14], $PbSnO_3$–$Pb(Mg_{1/3}Nb_{2/3})O_3$–$PbTiO_3$ [15], $BiScO_3$-$PbTiO_3$-$Pb(Cd_{1/3}Nb_{2/3})O_3$ [16], and even present new methods to induce MPB into the ordinary ferroelectric ceramics e.g., $BaTiO_3$ [17].

Recently, Park *et al*. [18] and Hyun *et al*. [19] investigated the dielectric properties of the Zr-doped $HfO_2$ ($Hf_{1-x}Zr_xO_2$ (HZO)) thin films. Their study proposed the existence of MPB phenomenon in $Hf_{0.5}Zr_{0.5}O_2$ composition. On one end (x = 0), it is monoclinic (m-phase) and thus paraelectric, whereas on the other end (x = 1), it is tetragonal (t-phase) and purely anti-ferroelectric (AFE). In between, by adjusting the fabrication conditions, it exhibits a non-centrosymmetric orthorhombic ferroelectric (FE) phase (o-phase) [20]. The multiphase in HZO is located between the boundary of the FE and AFE phases [18-20]. The existence of MPB in HZO system opened a new challenge to increase the dielectric properties of a silicon-compatible compound because the use of $ZrO_2$ and $HfO_2$ in the semiconductor fabrication process is in the mature stage.

The phase evolution in HZO is largely affected by its composition, film thickness, and other thin film processing parameters, and therefore, the region of multiphase formation is influenced by the fabrication conditions [21-33]. In contrast to other high-$\epsilon_r$ dielectrics, where the $\epsilon_r$ value decreases with decreasing film thickness, these films show increasing $\epsilon_r$ values with decreasing film thicknesses in the $\sim$ 5−20 nm range by adjusting the MPB region [18], which is highly promising for future dynamic random access memories. Therefore, this approach emerges as an alternative method for engineering an extremely thin equivalent oxide thickness (EOT).

Despite the numerous advantages of the existence of the MPB region in HZO films, very few studies have been conducted to date, and there has not been any direct evidence that supports the presence of the MPB phenomenon in HZO materials [18-20]. In this study, we fabricate the HZO-based Metal-Insulator-Metal (MIM) capacitors with different $\frac{t-phase}{o-phase}$ ratios ( $f_{\frac{t}{o}}$) to investigate the possible existence of the MPB region by evaluating the effect of $f_{\frac{t}{o}}$ on $\epsilon_r$. This study provides evidence of the existence of the MPB region in HZO films.

The thermal expansion coefficient (TEC) of capping electrodes appeared to be a crucial parameter in facilitating or impeding the formation of o-phase in $HfO_2$-based thin films, thus affecting $f_{\frac{t}{o}}$. It was established that metals with relatively low TEC compared to HZO provided additional driving force for t- to o-phase transition during rapid thermal annealing (RTA) [34]. Moreover, suppression of twin deformation through the formation of m-phase can be substantially prevented [35]. Another parameter that can affect $f_{\frac{t}{o}}$ is the ozone dosage during the atomic layer deposition (ALD) of $HfO_2$-based thin film. The primary role of the ozone $O_3$ pulses is to infuse oxygen into Hf and Zr layers, leading to the formation of $HfO_2$ and $ZrO_2$ compounds, respectively. Moreover, the removal of C−O bonds from the sample by applying $O_3$ pulses can remarkably change the subsequent annealing behavior of the as-grown film. The C–O bonds can remain between pure $HfO_2$ domains owing to the incomplete chemical reaction of the precursors that subsequently prevent the agglomeration of the nanoscale domains, resulting in the stabilization of the tetragonal phase and thus change of the $f_{\frac{t}{o}}$ [35-38]. It has been proven that the ratio of tetragonal to orthorhombic phases $f_{\frac{t}{o}}$ can be adjusted by applying different ozone dosages during deposition [35]. Therefore, fabrication of HZO films at different ozone pulses and adjusting annealing conditions enabled us to obtain MIM stacks with different $f_{\frac{t}{o}}$. The evaluation of $\epsilon_r$ of the MIM stacks with different $f_{\frac{t}{o}}$ is a promising approach to investigate the possible existence of the MPB region in HZO-based devices.

## Experiments

The ~10-nm HZO films were deposited on 50-nm thick W bottom electrodes sputtered on $SiO_2$/Si substrate using the ALD technique. Different ozone pulse durations from 2 to 30 s were applied during each deposition. The substrate temperature was maintained at 250 ºC. The Hf [N–($C_2H_5$) $CH_3$]$_4$ and Zr [N–($C_2H_5$) $CH_3$]$_4$ precursors (EG Chem CO.) were used as the Hf and Zr metal sources, respectively. The growth rate of $HfO_2$ and $ZrO_2$ were almost the same (~1 Å/cycle). After deposition, all films were capped with a 50-nm W electrode using rf-sputtering technique to fabricate MIM capacitors. Finally, W/HZO/W capacitors with different electrode areas were passed through an annealing process in ambient $N_2$ under two different conditions, i.e. 500 ºC for 30 s (A1) and 700 ºC for 5 s (A2), to achieve different t-/o-phase ratios $f_{\frac{t}{o}}$ according to these references [35, 39].

The elemental composition was evaluated through X-ray photoelectron spectroscopy (XPS). A high-resolution transmission electron microscopy (HRTEM) and an X-ray reflectometer (XRR) were used to measure the film thickness. We have considered the oscillations in the XRR pattern and using equation (1) the film thickness was obtained.

$$t \sim \frac{\lambda}{2} \frac{1}{\theta_{m+1} - \theta_m} \qquad (1)$$

Where λ is wavelength of X-ray, $\theta_{m+1}$ and $\theta_m$ are the position of *(m+1)-th* and *m-th* interference maximums, respectively.

The crystalline structures of the films were investigated using an X-ray diffractometer (XRD) within a grazing incidence. The $f_{\frac{t}{o}}$ was evaluated through a double-peak fitting of the o/t-phase characteristic Bragg peak. The ferroelectric properties of MIM capacitors were measured using a LC*II* ferroelectric precision tester (Radiant Technologies), and the dielectric permittivity was extracted from the small signal (50 mV) measurement of capacitance versus electric field at 10 kHz using the Keysight B1500A semiconductor device parameter analyzer. All measurements were performed at room temperature.

## Results and discussion

Figure 1 shows the XPS data of the as-deposited sample that revealed the Zr/Hf ratio to be approximately 1, indicating the $Zr_{0.5}Hf_{0.5}O_x$ composition which is a promising system to demonstrate the MPB phenomenon, as investigated by Park *et al*. [18]. The composition of deposited films was estimated by XPS intensities of the main peaks of Hf 4f, Zr 3d and O 1s.

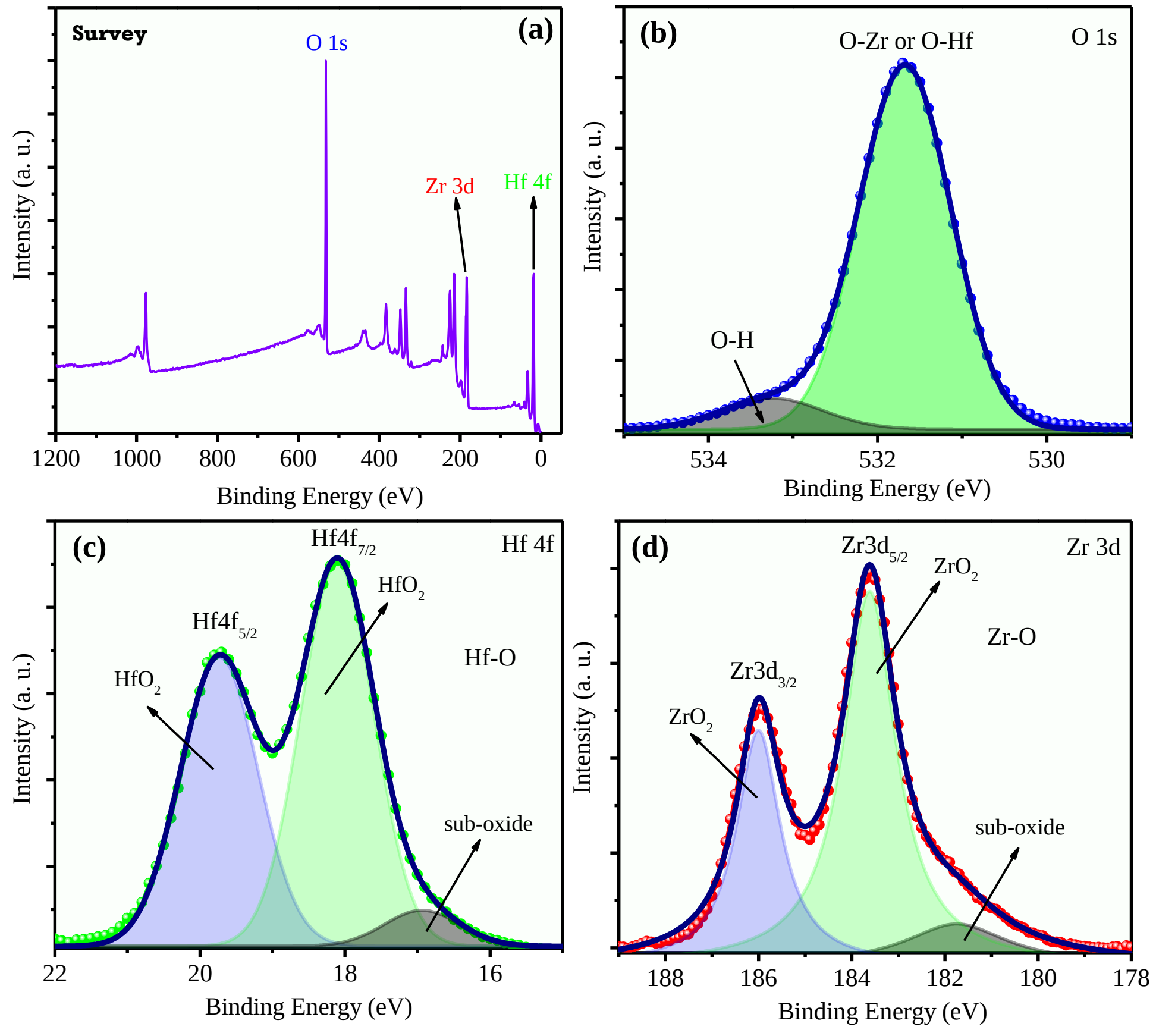

**Figure 1. (a)** X-ray Photoelectron Spectroscopy (XPS) survey spectrum of the as – deposited $Hf_{0.5}Zr_{0.5}O_2$ sample. (b) O 1s, (c) Hf 4f and (d) Zr 3d XPS spectra.

The atomic ratio of Zr, Hf and O acquired from the calculation of the ratio of respective area under the Zr 3d, Hf 4f and O 1s peaks considering atomic sensitivity factors. Our investigation shows that the ozone dosage or annealing conditions do not affect the Zr/Hf ratio and it is almost the same for all four samples.

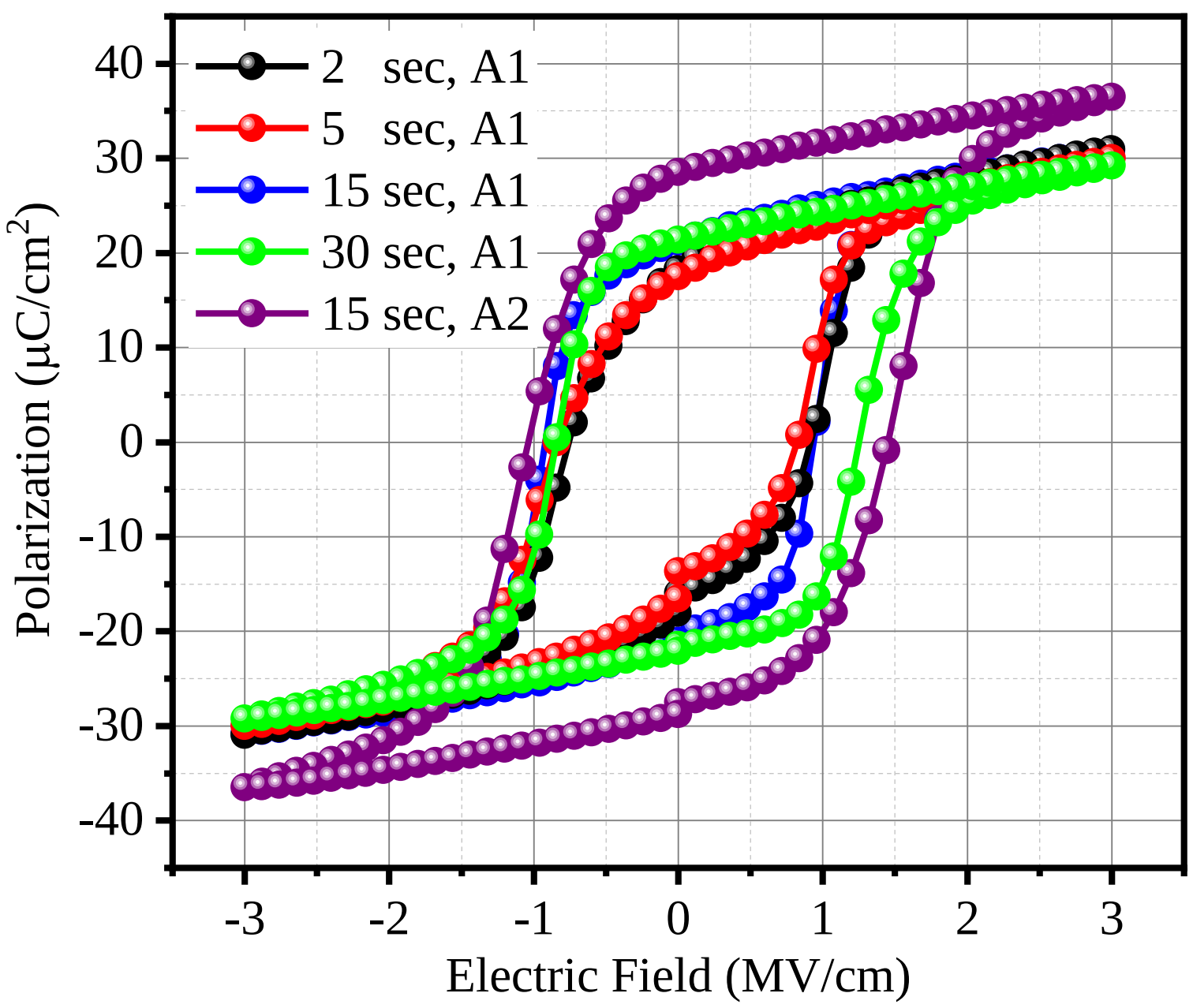

**Figure 2.** P–E curves of HZO devices fabricated under different conditions. A1 presents annealing at 500 ºC for 30 min, whereas A2 presents annealing at 700 ºC for 5 sec.

The pristine polarization versus electric field (P–E) curves for the MIM devices fabricated under varying conditions are presented in Figure 2. The increase in ozone dosage during the film deposition resulted in a more open loop that might be due to the suppression of the mechanisms that cause the *wake-up effect* [35]. Our recent study presented in reference [35] showed that the carbon contaminants, which stabilize the t-phase, can be removed through ozone treatment. Meanwhile, a higher ozone dosage can provide sufficient oxygen to form stoichiometric HZO films. Therefore, a de-pinched P–E loop was expected as the ozone dosage increased during the deposition. The sample annealed at 700 °C for 5 s (A2), resulting in a very high $2P_r$ value of ~ 63 μC/cm$^2$. This huge increase in $2P_r$ value compared to the samples annealed at 500 °C (A1) was attributed to the high mechanical stress applied to the HZO film during the cooling step of RTA, according to equation 2.

$$\epsilon_{\mathrm{T}}(T_A) = \int_{RT}^{T_{\mathrm{A}}} (\alpha_{\mathrm{film}} - \alpha_{\mathrm{Cap}})\, dT \qquad (2)$$

where $\alpha_{\mathrm{film}}$ and $\alpha_{\mathrm{Cap}}$ are the TEC of the film and capping materials, respectively, and $T_A$ is the post-annealing temperature. The TEC of W is lower than HZO film, which causes a tensile strain on HZO film during the cooling step of RTA. The tensile strain on HZO can facilitate the formation of o-phase by suppression of the twin deformation, which causes m-phase formation.

Figure 3a presents HRTEM image of the sample annealed at 700 °C for 5 sec. From the XRR scan (Figure 3b) the thickness was calculated using equation 1. An almost 10 nm film was deposited on W bottom electrode during the ALD process, which was confirmed by the cross-sectional HRTEM image. XRD results presented in figures 3c and 3d, indicate the two important features of different devices. In all MIM stacks, the formation of the m-phase was suppressed, and only t- and o-phase characteristic Bragg peaks were observed at approximately $2\theta$ = 30.5º. The increase in ozone dosage is accompanied by a left-shift of the t/o characteristic Bragg peaks (Figure 3d). This left-shift revealed an important structural feature of different HZO films. The fraction of the non-centrosymmetric o-phase in HZO films increases as ozone dosage increases. Therefore, the $2P_r$ value increased as illustrated in figure 2. Consequently, the ozone dosage during the film deposition can adjust $f_{\frac{t}{o}}$. Annealing at higher temperatures (700 ºC) further shifted the t/o characteristic Bragg peak to the left side and almost suppressed the formation of the t-phase (therefore, $f_{\frac{t}{o}} \sim 0$) after the cooling process to room temperature. This caused a huge increase in the $2P_r$ value as was already presented in figure 2.

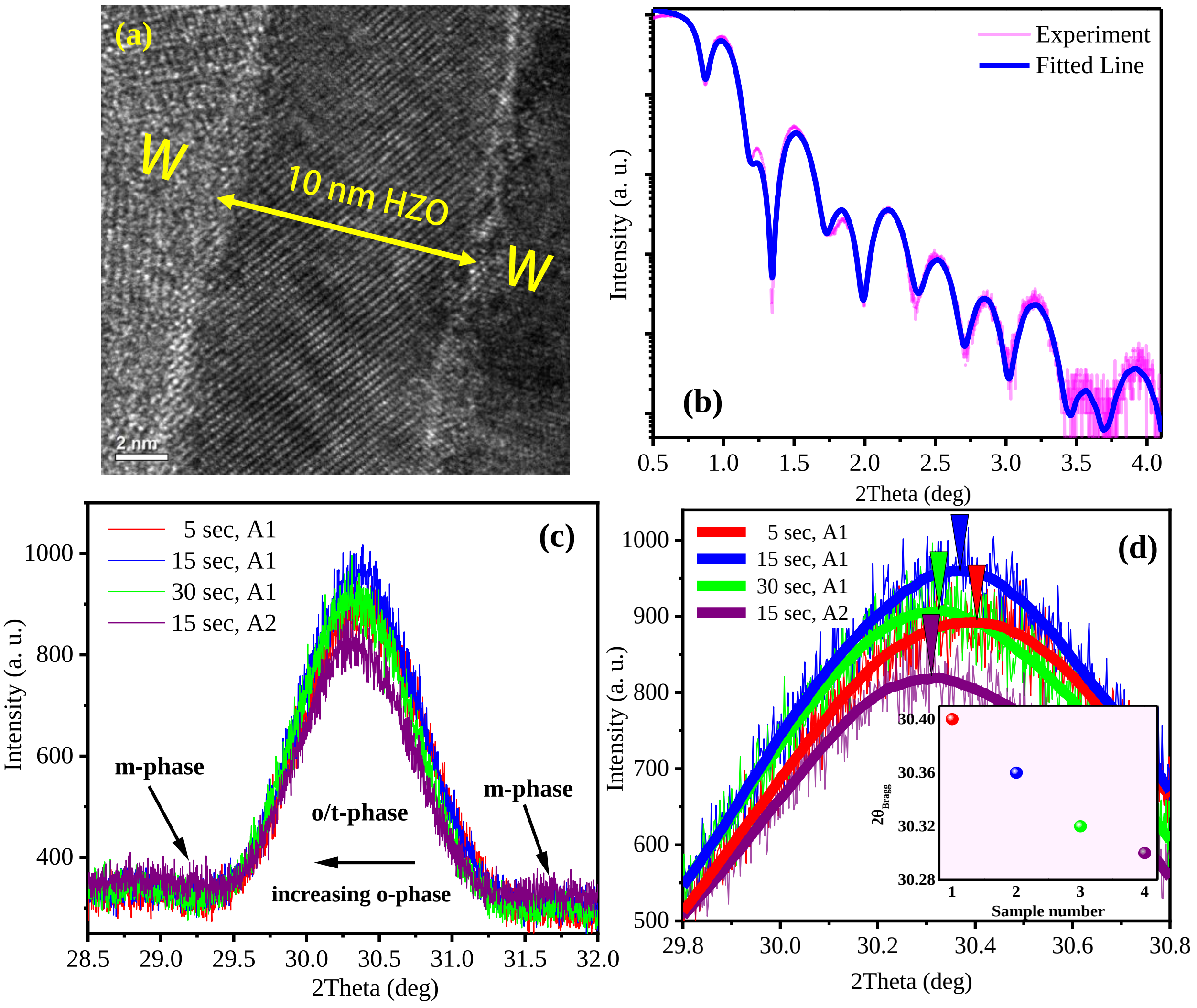

**Figure 3.** (a) A cross-sectional HRTEM of W/HZO/W device. (b) X-ray reflectivity of the ~10-nm films grown at 5 sec ozone dosage. (b) XRD patterns for HZO films deposited under different conditions. (d) The magnified region of XRD patterns showing the evolution of o/t Bragg peak versus fabrication conditions.

To evaluate the structural changes of HZO film fabricated at different conditions, the characteristic o/t peaks that were detected around 2θ ~ 30.5º was precisely investigated. Figure 4 and table 1 revealed that the $f_{\frac{t}{o}}$ substantially changes depends on the fabrication condition. The $f_{\frac{t}{o}}$ values were directly calculated by considering the relative area below each characteristic peak. Therefore, different fabrication procedures cause the deposition of HZO films with different $f_{\frac{t}{o}}$ (Table 1).

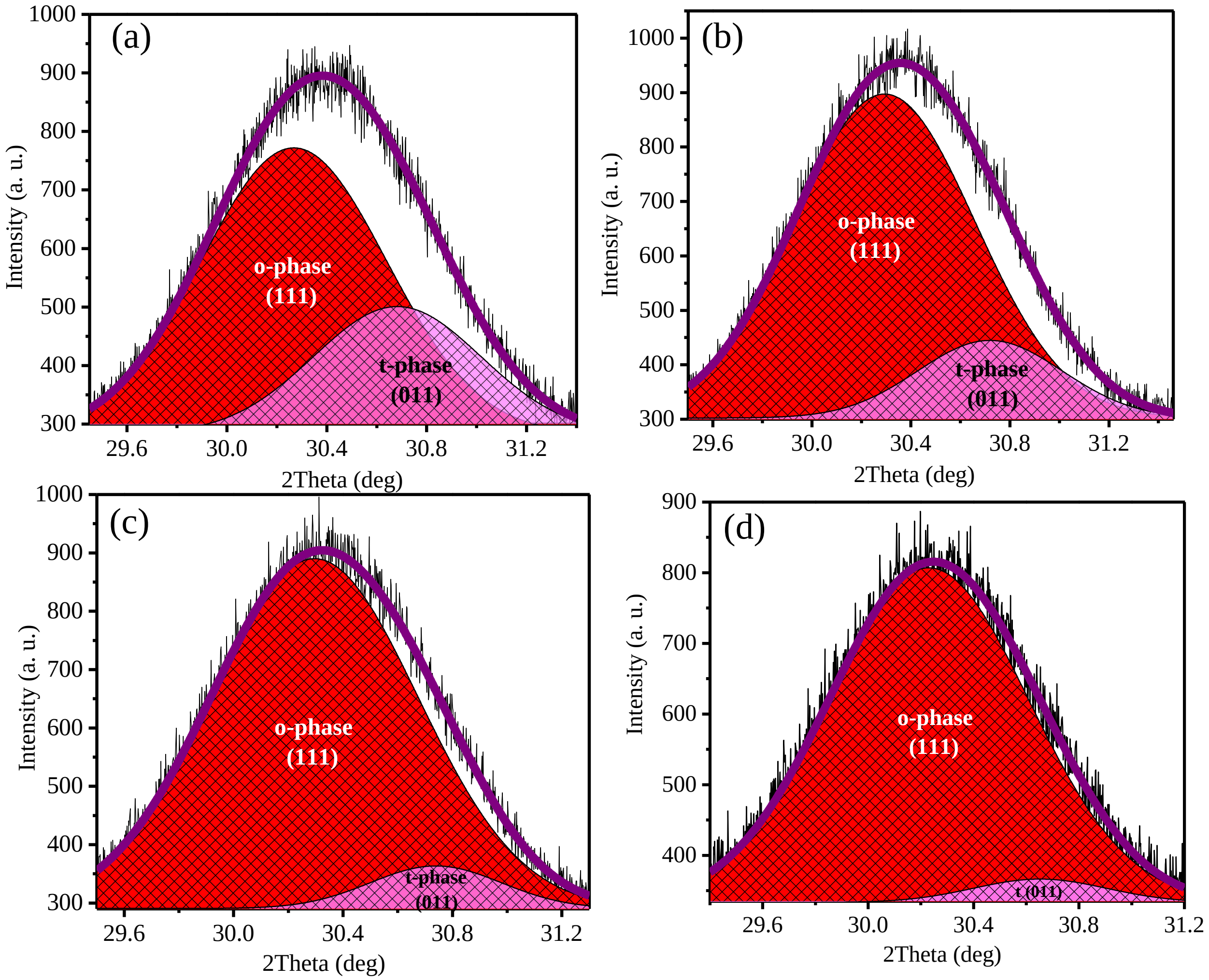

**Figure 4.** X-ray diffraction patterns of the HZO films fabricated at (a) 5 s ozone dosage A1, (b) 15 s ozone dosage A1, (c) 30 s ozone dosage A1, and (d) 15 s ozone dosage A2. The red area presents o (111) peak and the pink area presents t (011) peak.

**Table 1.** The calculated $f_{\frac{t}{o}}$ in HZO films deposited at different ozone dosages and annealing conditions. The $f_{\frac{t}{o}}$ was directly evaluated using the relative area below characteristic Bragg peaks from figure 4.

| Fabrication condition | 5 s $O_3$, A1 | 15 s $O_3$, A1 | 30 s $O_3$, A1 | 15 s $O_3$, A2 |
|---|---|---|---|---|
| $f_{\frac{t}{o}}$ | 0.41 | 0.19 | 0.08 | 0.04 |

Recently, Park *et al.* [18, 19] investigated the existence of MPB in HZO thin films. According to their study, the $Hf_{0.5}Zr_{0.5}O_2$ film is the promising candidate to show the MPB region. However, different studies showed that the compositional region ($x$) in which the $Hf_{1-x}Zr_xO_2$ thin films show MPB-like behavior is greatly affected by the fabrication process, as the coexistence of the t/o-phase can be substantially altered during the fabrication process.

It has been shown that materials exhibit maximum piezoelectric and dielectric responses near the MPB [5]. The response to external stimuli is significant near the MPB, as several phases (different phases have different charge responses, such as dielectric, ferroelectric, and anti-ferroelectric) co-exist and there is nearly no energy barrier separating them. Consequently, a slight mechanical or electrical perturbation could induce a significant charge response. Therefore, we measured the $\epsilon_r$ values as a function of applied bias electric field for different HZO films (Figure 5a) and established that the sample containing higher $f_{\frac{t}{o}}$ shows higher $\epsilon_r$ in different ranges of applied bias voltage. The butterfly-like capacitance versus bias voltage is due to the increase in motion of domain walls under the application of a strong DC electric field. It has been shown that the application of strong electric field on HZO film can induce the phase transition from t- to o-phase affecting the MPB region [40]. Therefore, apart from the domain wall permittivity, the capacitance of the HZO-based MIM stack can be further increased by the application of a strong electric field that causes the phase transition between t- and o-phases.

The sample with the minimum $f_{\frac{t}{o}}$, which showed strong ferroelectric polarization (Figure 2), demonstrates the lowest electric field-induced permittivity compared to the other samples. Conversely, the sample with the maximum $f_{\frac{t}{o}}$ (confirmed through XRD measurement) shows the highest electric field-induced permittivity. Therefore, the $f_{\frac{t}{o}}$ in the HZO film determines its dielectric permittivity under a bias electric field. A higher fraction of the t-phase results in a higher response from the film under external stimuli (electric field). According to the MPB-based point of view, the highest ratio of o/t-phase ($f_{\frac{t}{o}}$) provides the highest density of the MPB region in the deposited materials. Therefore, an external stimulus such as a strong electric field receives strong feedback from the material. Figure 5b shows $\epsilon_r^{max}$ (Maximum dielectric constant) and $\epsilon_r^{E=0}$ (Dielectric constant at zero-biased field) versus $f_{\frac{t}{o}}$ for different samples. Two important features can be observed in this figure. First, as $f_{\frac{t}{o}}$ increases, both dielectric constants ($\epsilon_r^{max}$ and $\epsilon_r^{E=0}$) increased and the sample with the maximum $f_{\frac{t}{o}}$ attained the maximum $\epsilon_r$. The second feature that is a meaningful trend, is the difference between two dielectric constants *i.e.*, $\Delta\epsilon_r = \epsilon_r^{max} - \epsilon_r^{E=0}$ for each sample (Inset of figure 5b). This difference indicates the extent to which the biased electric field can increase the $\epsilon_r$ in each sample. The sample with the highest $f_{\frac{t}{o}}$ has the highest electric field-induced increase in dielectric constant ($\Delta\epsilon_r$ ), whereas the sample with the lowest $f_{\frac{t}{o}}$ has the minimum increase in dielectric constant under the application of a biased field. The difference between $\epsilon_r^{max}$ and $\epsilon_r^{E=0}$ ($\Delta\epsilon_r$ ) increases as $f_{\frac{t}{o}}$ increases.

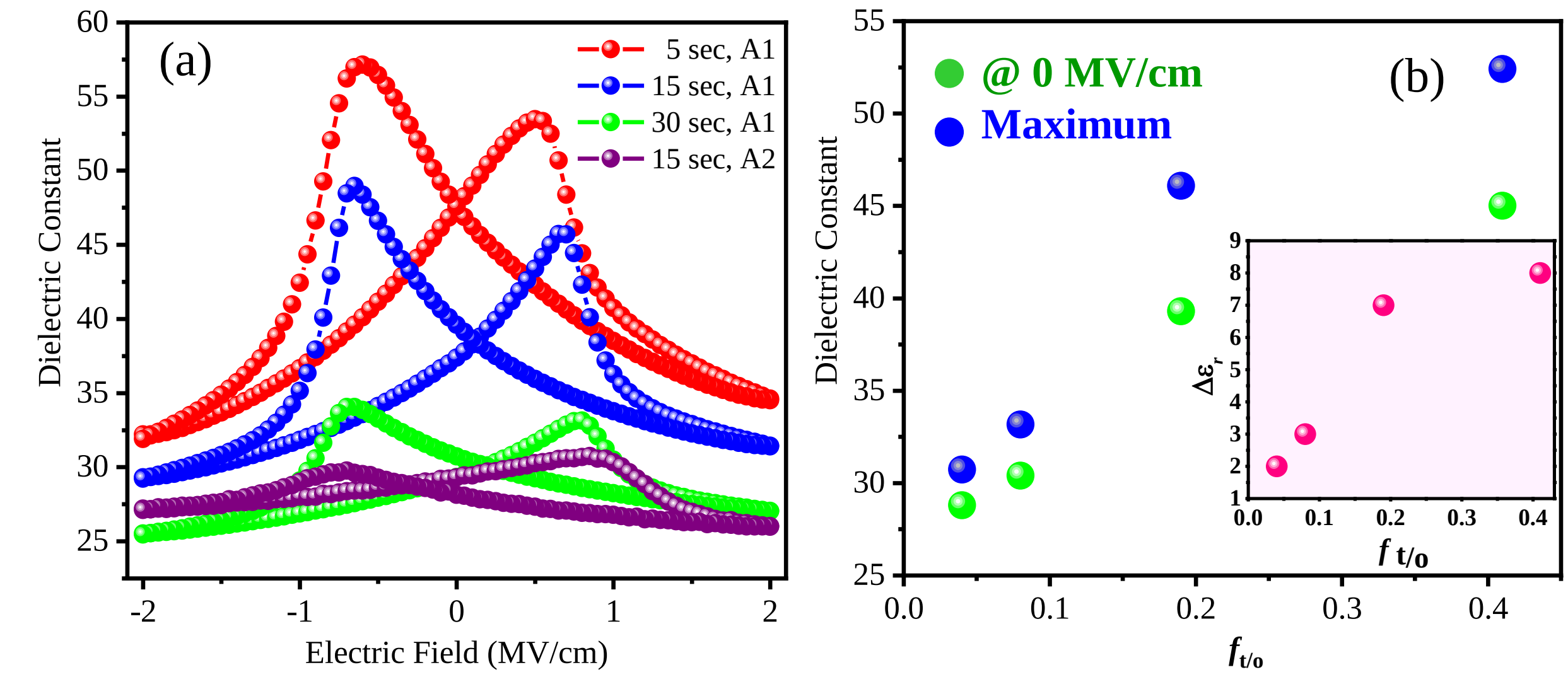

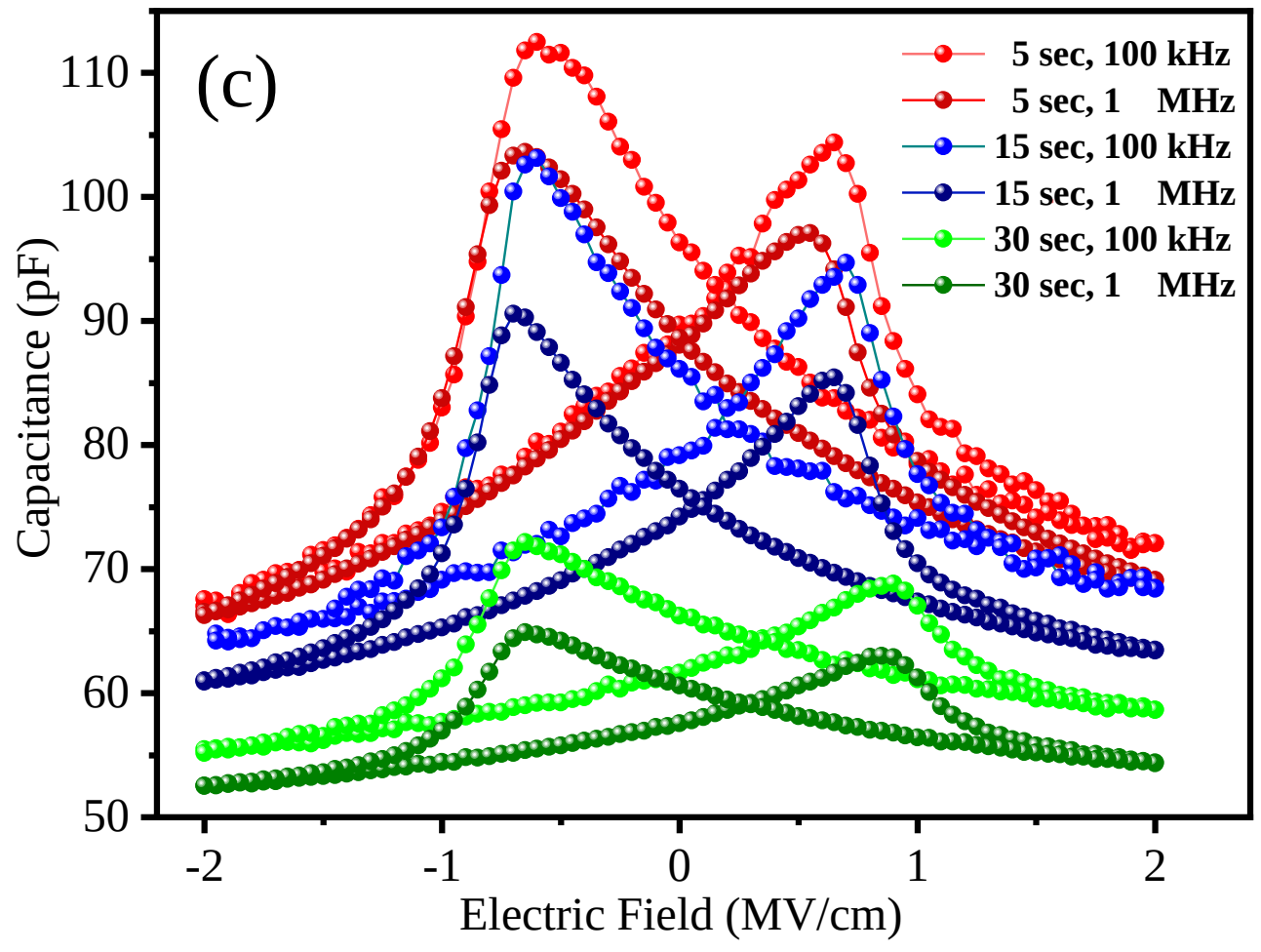

**Figure 5.** (a) Dielectric constant vs. bias electric field for the samples grown under different conditions measured at 10 kHz. (b) The $\epsilon_r^{max}$ and $\epsilon_r^{E=0}$ versus $f_{\frac{t}{o}}$. The inset shows the difference between $\epsilon_r^{max}$ and $\epsilon_r^{E=0}$ ($\Delta\epsilon_r$ ) as a function of $f_{\frac{t}{o}}$. (c) Capacitance vs. electric field for different samples as a function of applied frequency. The Red branch corresponds to the samples grown at 5 s ozone dosage, Blue branch at 15 s, and Green branch at 30 s. All samples were annealed under the A1 conditions.

To further investigate the C–E behavior of MIM devices fabricated under different conditions, the capacitance was measured at higher frequencies, i.e., 100 kHz and 1 MHz. The results for different capacitors show a decrease in dielectric capacitance as the frequency increases (Figure 5c). This behavior was expected as some polarization mechanisms can be suppressed at higher frequencies, in particular, space-charge polarization that occurs owing to the structural defects [41, 42]. Therefore, by increasing the frequency, we attempted to separate the contribution of space-charge polarization from the intrinsic polarization mechanisms (i.e., electronic, ionic, dipole, and domain wall polarizations). Figure 5c shows that even at 1 MHz, the sample with the maximum $f_{\frac{t}{o}}$ (Red branch) exhibits the highest capacitance compared to the other samples. These results provide evidence of the existence of the MPB region due to the coexistence of t-and o-phases in HZO films.

## Conclusion

The existence of the MPB region in HZO thin films was investigated through XPS, XRD, and dielectric measurements. The HZO thin films with different $f_{\frac{t}{o}}$ were deposited through adjusting fabrication conditions. The sample with the lowest $f_{\frac{t}{o}}$ showed the smallest $\epsilon_r$, whereas that with the maximum $f_{\frac{t}{o}}$ showed a large $\epsilon_r$ of ~55. There has been a direct relation between $f_{\frac{t}{o}}$ and $\epsilon_r$. Moreover, the $f_{\frac{t}{o}}$ determined the electric field-induced increase of $\epsilon_r$ in each sample. These observations provide strong evidence supporting the existence of the MPB region in HZO films.

## Acknowledgements

This work was supported by the National Research Foundation of Korea funded by the Korea government (MSIT), Grant No. NRF-2018R1A3B1052693. The authors thank Dr. Stanislav Kamba, Dr. Writam Banerjee and Dr. Nikam for helpful discussions.